\documentclass[%
 prl,superscriptaddress,
 amsmath,amssymb,
 aps,
twocolumn, % https://tex.stackexchange.com/questions/160730/putting-two-column-wide-figure-in-revtex4-1
longbibliography]{revtex4-2}
\usepackage[utf8]{inputenc}
\usepackage{graphicx}% Include figure files
\usepackage{dcolumn}% Align table columns on decimal point
\usepackage{bm}% bold math
\usepackage{physics}
\usepackage{xcolor}
\usepackage{lipsum}
\usepackage[export]{adjustbox}

\makeatletter
\renewcommand*{\@fnsymbol}[1]{\ensuremath{\ifcase#1\or \dagger\or *\or \ddagger\or
   \mathsection\or \mathparagraph\or \|\or **\or \dagger\dagger
   \or \ddagger\ddagger \else\@ctrerr\fi}}
\makeatother

\usepackage{verbatim} % for block commenting

\begin{document}

\preprint{APS/123-QED}

\title{Robust Quantum Control via Multipath Interference for Thousandfold Phase Amplification in a Resonant Atom Interferometer}
\author{Yiping Wang\@{*}}
\affiliation{Department of Physics and Astronomy and Center for Fundamental Physics, Northwestern University}

\author{Jonah Glick\@{*}}
\affiliation{Department of Physics and Astronomy and Center for Fundamental Physics, Northwestern University}

\author{Tejas Deshpande\@{*}}
\affiliation{Department of Physics and Astronomy and Center for Fundamental Physics, Northwestern University}

\author{Kenneth DeRose\@{*}}
\affiliation{Department of Physics and Astronomy and Center for Fundamental Physics, Northwestern University}

\author{Sharika Saraf}
\affiliation{Department of Physics and Astronomy and Center for Fundamental Physics, Northwestern University}

\author{Natasha Sachdeva}
\affiliation{Department of Physics and Astronomy and Center for Fundamental Physics, Northwestern University}
\affiliation{Q-CTRL, Quantum Applications and Algorithms Division}

\author{Kefeng Jiang}
\affiliation{Department of Physics and Astronomy and Center for Fundamental Physics, Northwestern University}

\author{Zilin Chen}
\affiliation{Department of Physics and Astronomy and Center for Fundamental Physics, Northwestern University}

\author{Tim Kovachy}
\email{timothy.kovachy@northwestern.edu}
\affiliation{Department of Physics and Astronomy and Center for Fundamental Physics, Northwestern University}

\begin{abstract} 

 We introduce a novel technique for enhancing the robustness of light-pulse atom interferometers against the pulse infidelities that typically limit their sensitivities. The technique uses quantum optimal control to favorably harness the multipath interference of the stray trajectories produced by imperfect atom-optics operations. We apply this method to a resonant atom interferometer and achieve thousand-fold phase amplification, representing a fifty-fold improvement over the performance observed without optimized control. Moreover, we find that spurious interference can arise from the interplay of spontaneous emission and many-pulse sequences and demonstrate optimization strategies to mitigate this effect. Given the ubiquity of spontaneous emission in quantum systems, these results may be valuable for improving the performance of a diverse array of quantum sensors. We anticipate our findings will significantly benefit the performance of matter-wave interferometers for a variety of applications, including dark matter, dark energy, and gravitational wave detection.

\end{abstract}

\maketitle

\begin{figure}
\includegraphics[width=3.25in]{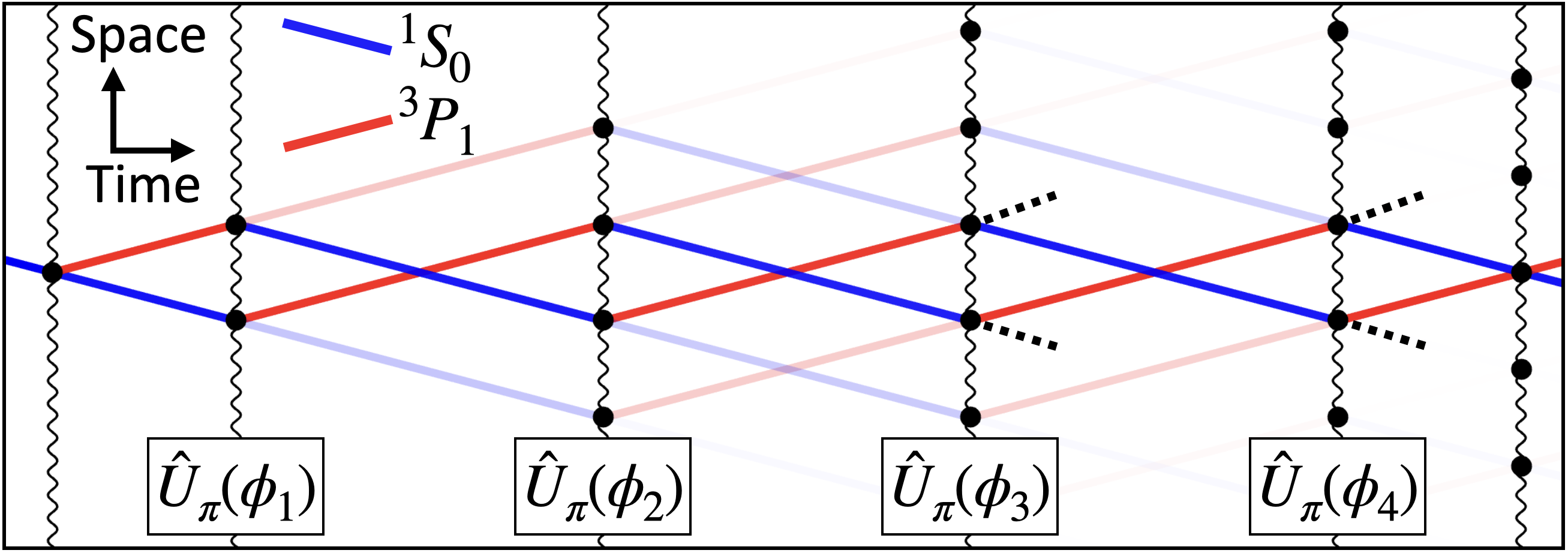}
\caption{\label{fig:resonant_ai_schematic} Resonant light-pulse atom interferometry: Mirror ($\pi$) and beamsplitter pulses split, redirect, and interfere interferometer arms (atom-laser interactions at black dots). Imperfect mirror operations $\hat{U}_{\pi}(\phi_i)$ create stray paths. The laser phases for the mirror pulses ($\phi_i$ for the $i$th pulse) are control parameters tuned via quantum optimal control to limit the spatial spread of trajectories, leveraging intra-sequence interference between trajectories. Dashed black lines indicate trajectories suppressed through destructive interference. Greater opacity of the atom paths (blue and red) denotes larger population.
}
\end{figure}

Quantum sensing is a powerful tool for both fundamental physics and practical measurements \cite{chou2023quantum,Safronova2018,degen2017quantum}, including the detection of oscillating signals \cite{chou2023quantum,Safronova2018,degen2017quantum,antypas2022new,Dimopoulos2008_GW,hollberg2017optical,banerjee2023phenomenology,Arvanitaki2018_darkmatter,Graham2016_darkmatter,Graham2018axion,badurina2022refined,yu2011gravitational,graham2013new,abe2021matter,abou2020aedge,abend2023terrestrial,ZAIGA2020, badurina_2020,canuel2018exploring, chiow2018multiloop,coslovsky2017ac}. Resonant \cite{kafle2011analysis,pedernales2020motional,Horikoshi2017,burke2008confinement,moukouri2021multi,pushin2009decoherence} light-pulse atom interferometers \cite{dubetsky2006atom,schubert2021multi,jaffe2018efficient, marzlin1996state,mcguirk2002sensitive,sidorenkov2020tailoring,chiow2018multiloop,kim2022one,di2024optimal,coslovsky2017ac,Graham2016_GW}, a class of light-pulse atom interferometers \cite{tino2014atom,cronin2009optics,bongs2019taking,narducci2022advances}, employ sequential mirror pulses to periodically reverse the directions of the interferometer arms, creating loops that repeat at a tunable target frequency (see Fig. \ref{fig:resonant_ai_schematic}). These resonant (or multiloop) atom interferometers can be used as highly sensitive detectors for various oscillatory signals \cite{Graham2016_GW,torres2023detecting,abou2020aedge,baum2024gravitational,schubert2019scalable,Arvanitaki2018_darkmatter,Graham2016_darkmatter,Graham2018axion,abou2020aedge,abend2023terrestrial,di2024optimal,chiow2018multiloop,Graham2016_darkmatter,coslovsky2017ac,jaffe2018efficient}. 

Oscillatory signals, whether intrinsic or engineered through experimental design, span a wide range of applications within atom interferometry, making resonant atom interferometers highly versatile for many of the field’s current research objectives. The target frequency at which the interferometer is maximally sensitive is easily tunable by adjusting the temporal spacing between mirror pulses, with the upper frequency limit constrained by the pulse repetition speed, which is in turn determined by the available Rabi frequency.  Increasing the number of loops enhances sensitivity at the target frequency while suppressing off-resonant noise \cite{Graham2016_GW,dubetsky2006atom,abe2021matter}, akin to applying lock-in detection \cite{kotler2011single,boss2017quantum,maze2008nanoscale,de2011single,shaniv2017quantum,schmitt2017submillihertz,shibata2021quantum,zhuang2021many,kolkowitz2012coherent} and dynamical decoupling \cite{viola1999dynamical,GULLION1990,yang2011preserving,lidar2014review,biercuk2009optimized,Genov_2017} to matter waves.

Multiloop atom interferometers are powerful tools for probing fundamental physics.
One major application is gravitational wave detection in the frequency range of 0.03 - 3 Hz, which complements the sensitivity of laser interferometers like LIGO and LISA \cite{abou2020aedge,abend2023terrestrial}. Large-scale atom interferometric detectors are being developed worldwide by an interdisciplinary community \cite{abou2020aedge,abend2023terrestrial,abe2021matter,badurina_2020,Hartwig2015,Asenbaum2020_equivalence,overstreet2022observation,canuel2018exploring,ZAIGA2020}. Resonant atom interferometry has been proposed to optimize the sensitivity of these interferometers to gravitational waves \cite{Graham2016_GW,torres2023detecting,abou2020aedge,baum2024gravitational,schubert2019scalable}. Another application is searches for wavelike dark matter candidates, which are considered among the most promising dark matter models \cite{antypas2022new}. Signatures of these candidates would manifest as  an oscillatory signal in an atom interferometer, and resonant atom interferometers are expected to significantly enhance sensitivity across a broad frequency
range from 10 mHz to several MHz \cite{Arvanitaki2018_darkmatter,Graham2016_darkmatter,Graham2018axion,abou2020aedge,abend2023terrestrial,di2024optimal,jaffe2018efficient}.

Resonant atom interferometers also have practical applications. These interferometers could be used in geophysical studies to measure oscillating seismically-induced accelerations \cite{coslovsky2017ac}, either from direct vibrations of the Earth or from density fluctuations caused by seismic waves \cite{canuel2018exploring}. The latter measurements are not only of intrinsic geophysical interest \cite{canuel2018exploring}, but are also crucial for characterizing noise backgrounds in both laser- and atom-based gravitational wave detectors \cite{canuel2018exploring,mitchell2022magis,badurina2023ultralight,harms2015terrestrial}. Multiloop interferometers are also useful in inertial navigation because they act as selective sensors for rotations or accelerations \cite{dubetsky2006atom,sidorenkov2020tailoring} and have the potential to amplify signals in atomic gyroscopes \cite{kim2022one,schubert2021multi}.

In some cases, experiments can be designed to cause otherwise DC signals to manifest as oscillatory in order to enable the signal amplification and noise suppression benefits of resonant detection. One example is an ongoing dark energy search experiment \cite{chiow2018multiloop}, which employs a spatially modulated source mass to induce a modulation in the dark energy field in the frame of the atoms.
A similar approach has been suggested for measuring the Newtonian gravitational constant \cite{coslovsky2017ac}. The improvement to resonant atom interferometry presented in our work may encourage the further development of experiments which render DC signals oscillatory.

A critical challenge impeding the full metrological potential of resonant interferometers is the adverse impact of mirror pulse infidelities, which limit the total number of loops that can be performed \cite{coslovsky2017ac,jaffe2018efficient,kim2022one}. With imperfect pulse efficiency, each atom's wavefunction diverges into many `stray' paths that deviate from the central two interferometer arms (Fig. \ref{fig:resonant_ai_schematic}), thereby diminishing the population contributing to the signal and/or introducing deleterious effects through multipath interference \cite{stockton2011absolute,sidorenkov2020tailoring}\footnote{Note that in certain scenarios, multipath interference \cite{rahman2024bloch,deng1999temporal,moore1995atom} has been used to destructively reduce losses in other types of atom interferometers \cite{robert2010quantum,beguin2023atom}}.

Experimental tradeoffs and imperfections inevitably result in mirror pulse infidelities.  Two experimental parameters relevant to such tradeoffs are the temperature of the atom cloud and the ratio of the interferometer laser beam waist to the atom cloud size, noting that practical limitations exist on how small the atom cloud can be made \cite{abend2023terrestrial, loriani2019atomic, ammann1997delta, chu1986proposal}. 
Accepting a hotter cloud can simplify and expedite cloud preparation and enable larger atom numbers, but can lead to large thermal Doppler detuning spreads which degrade pulse fidelities \cite{abend2023terrestrial,mcguinness2012high}.
With typical constraints on the total interferometer beam power, decreasing the beam waist increases intensity inhomogeneity across the cloud---which adversely affects pulse fidelity---but advantageously increases the peak intensity \cite{abend2023terrestrial}.
A large intensity reduces the susceptibility of the pulse efficiency to Doppler detunings \cite{metcalf1999laser}, allows more pulses per unit time---key for the efficacy of clock atom interferometers which use single-photon transitions on narrow clock lines \cite{graham2013new,yu2011gravitational,Hu2017_clockinterferometry,hu2019sr,abe2021matter,rudolf2020_689LMT,wilkasonFloquet}, especially in parallel operation \cite{abend2023terrestrial,Graham2016_GW}---and enables reduced incoherent photon scattering for multi-photon Raman, Bragg \cite{tino2014atom,cronin2009optics,kim202040}, or clock \cite{carman2024collinear} atom optics. 
Such tradeoffs are especially pronounced for long-baseline interferometry \cite{abou2020aedge,abe2021matter,badurina_2020,abend2023terrestrial,Hartwig2015,Asenbaum2020_equivalence,overstreet2022observation,canuel2018exploring,ZAIGA2020} and portable sensors with hotter atom clouds \cite{bongs2019taking,narducci2022advances}.
A maximum of 400 loops has been achieved, but required preparing a Bose-Einstein condensate confined in a waveguide to attain single-pulse efficiencies of 99.4\% \cite{kim2022one}.
It is therefore essential to develop resonant atom interferometers where the signal remains robust against mirror pulse infidelities, directly mitigating the drawbacks associated with experimental tradeoffs.

Traditionally, individual beamsplitter or mirror operations in light-pulse atom interferometers \footnote{An exception involves the compensation of phase errors in a three-pulse Bragg Mach-Zehnder interferometer across the beam splitter and mirror pulses \cite{saywell2023enhancing}}, including large momentum effective atom optics operations composed of sequences of many pulses \cite{tino2014atom}, have been made robust using composite pulses \cite{Butts2013, dunning2014_composite, Berg2015_composite}, adiabatic rapid passage \cite{Kotru2015,jaffe2018efficient,kovachy2012adiabatic}, Floquet pulse engineering \cite{wilkasonFloquet} or acceleration \cite{rodzinka2024optimal}, quantum optimal control (QOC) \cite{saywell2018_QuOC_Mirror, saywell2020optimal, saywell_2020_biselective, goerz2021quantum, Goerz2023, saywell2023enhancing,saywell2023realizing,Chen2023,louie2023robust}, and coherent enhancement of Bragg pulse sequences \cite{beguin2023atom}. These methods usually, though not always \cite{beguin2023atom,rodzinka2024optimal}, require increased pulse area and/or duration, leading to increased signal degradation from incoherent photon scattering and/or reducing the number of pulses that can be applied per unit time. Moreover, despite enhancements from these methods, atom optics operations inevitably retain residual infidelities.

We instead introduce the method of making the entire interferometer sequence robust to imperfections in individual mirror operations by  leveraging QOC \cite{werschnik2007quantum,khaneja2005optimal,koch2022quantum,ansel2024introduction,magrini2021real,jandura2022time,tian2020quantum,poggiali2018optimal,rembold2020introduction,muller2022one} in a novel way to advantageously harness multipath interference between numerous stray trajectories. As a demonstration, we utilize this method to enhance the robustness of a resonant single-photon clock atom interferometer (using the 689\;nm $^1S_0-^3P_1(m_j=0)$ intercombination line of $^{88}$Sr)  to pulse infidelities, achieving 50-fold improvement in the resonant signal amplification---for a total of more than 500 loops---in spite of an individual pulse transfer efficiency of $\approx 90\%$, representative of what one might encounter for a long-baseline atom interferometer \cite{abend2023terrestrial} such as MAGIS-100 \cite{abe2021matter}. Our general approach is not constrained by the specific type of atom optics, and we anticipate that it will enable higher levels of precision across various atom interferometer designs, including those based on multi-photon transitions.

Moreover, we elucidate how the interplay between spontaneous emission and many-pulse sequences can produce spurious interference signals with unexpectedly high visibility, which can obscure the signals of interest. To address this, we present an optimization approach that effectively mitigates this effect. These findings could significantly enhance the performance of quantum sensors susceptible to spontaneous emission, especially as advancements in quantum control techniques enable increasingly complex pulse sequences \cite{nielsen2010quantum,agarwal2006quantum}. For instance, many-pulse sequences can improve the performance of atomic clocks used in gravitational wave and dark matter detection \cite{kolkowitz2016gravitational,zaheer2023quantum}, as well as optical spectrum analysis \cite{bishof2013optical}, and the effect we have identified will likely be crucial to consider in future applications of atomic clocks in these areas. Notably, we observe a substantial spurious interference signal using the same pulse phase sequence reported for the optical spectrum analyzer in \cite{bishof2013optical}, suggesting that similar effects may arise when extending this work to larger pulse numbers.

\begin{figure*}[hbt!]
\centering
\includegraphics[scale=1]{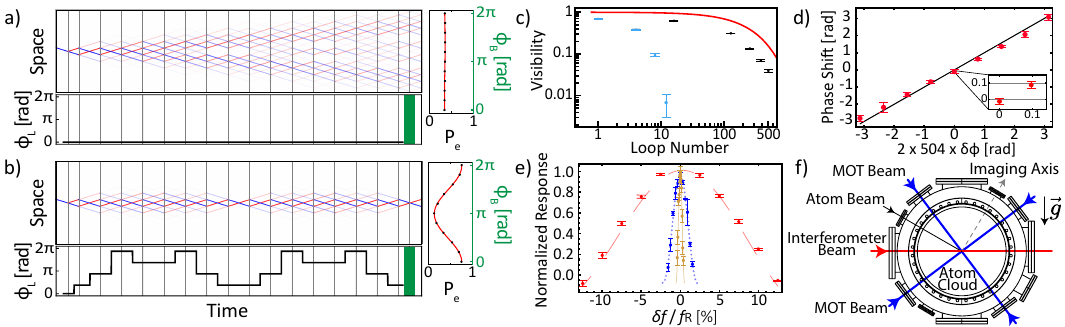}
\caption{\label{fig:rank3_introduction}
Introduction to the N=8-pulse sequence with optimized phases $(\frac{3\pi}{8},\frac{7\pi}{8},\frac{15\pi}{8},\frac{11\pi}{8},\frac{11\pi}{8},\frac{15\pi}{8},\frac{7\pi}{8},\frac{3\pi}{8})$. A comparison of the spatial spread of trajectories, the laser phases $\phi_L$, and experimentally collected interferometer fringes (excited state population $P_{e}$ vs. final beam splitter phase $\phi_{B}$) in 16-loop unoptimized (a) and optimized (b) sequences.
(c) Interferometer fringe visibility as a function of loop number for optimized (black) and unoptimized (light blue) sequences. The solid red curve indicates the spontaneous emission limit.  Error bars are generated via standard error from fits of the interferometer fringes.  (d) The phase amplification of the optimized sequence observed after 504 loops. Each data point represents an average derived from five fringe scans, with error bars calculated from the standard deviation. 
Inset shows the ability of the enhanced phase sensitivity to resolve bit discretization in the phase control for 504 loops. Gray lines indicate expected phase shifts for the minimum laser phase modulation, limited by bit discretization.
Average values and error bars, generated from standard error of the mean, are evaluated from 45 fringe scans. (e) Data of the self-normalized resonant response vs. offset ($\delta f$) from resonant frequency ($f_R$) with theory curves for the optimized sequence for loop numbers 8 (dashed red), 64 (dotted blue), and 256 (solid yellow). Error bars are generated via standard error from fits of the interferometer fringes. (f) The science chamber with the orientations of the atom beam, MOT beams, the interferometer beam, and the axis along which the imaging camera points (the `Imaging Axis’). A third MOT beam, along with push and fluorescence beams, come in and out of the page. The atom cloud is prepared in the center of the chamber, and the interferometer beam is perpendicular to the direction of gravity.
}
\end{figure*}

\textit{Open-loop optimization with QOC.} We use a semi-classical model \cite{hogan2008lightpulse, bongs2006high,antoine2003quantum} to compute the spread of trajectories in a multiloop interferometer under imperfect mirror operations for a particular choice of laser phases \cite{supplement}. We apply QOC to these interferometers by selecting the laser phases of the mirror pulses as control parameters \cite{saywell2020optimal,GULLION1990} and defining a cost function aimed at minimizing the spatial spread of trajectories throughout the interferometer cycle \cite{supplement}. Our approach leverages the fact that stray trajectories spawned by one $\pi$-pulse can spatially overlap and interfere with those spawned by previous $\pi$-pulses. The phase of the laser is imprinted on kicked trajectories, which we find can facilitate destructive interference among paths diverging from the central interferometer arms and constructive interference among stray paths that rejoin them, as visualized in Fig \ref{fig:rank3_introduction} (a-b).

We define an $N$-dimensional optimization problem where the phases of the first $N$ mirror pulses repeat $L/N$ times to compose an $L$-loop interferometer. We arrive at the sequence characterized in Figures \ref{fig:rank3_introduction} and \ref{fig:rank3_robustness} by tuning the phases of the mirror pulses to minimize the spread of trajectories over a set of $N=8$ laser phases repeated 8 times (see End Matter for further details). Our optimization approach not only yields sequences that prove effective experimentally, but also provides a visual explanation of why certain sequences outperform others.  For example, we notice that for the case of $L=8$, the universally robust (UR) UR-8 dynamical decoupling sequence, originally derived under an entirely different framework \cite{Genov_2017}, provides near-optimal results per our cost function. Following the metrics used in \cite{Genov_2017} would suggest that for $L$ loops, it is best to implement the corresponding UR-$L$ sequence. For larger $L$, however, we observe that applying UR-$L$ to a resonant atom interferometer leads to a reduced interferometer phase sensitivity to resonantly oscillating signals. The semi-classical model visually illustrates the reason for this: The further a trajectory strays from the central arms, the less phase information it carries about the resonantly oscillating signal due to the fewer imprinted laser phases. The UR-$L$ sequence leads to a spread of spatial trajectories which increases with $L$, thereby reducing overall interferometer sensitivity to oscillating signals as $L$ increases (see Fig. \ref{fig:nl_comparison} in End Matter). Applying the spatial spread cost optimization for $N=8$ at large $L$ produces pulse phases which happen to coincide with a repeated UR-8 pattern---outperforming UR-$L$ in our application---and optimizing over $N=16$ phases produces sequences distinct from the UR-$N$ formula owing to the reduced phase sensitivity associated with the UR-$N$ sequences at high $N$.
Extending our optimization from $N=8$ to $N=16$ enhances the interferometer visibility by 1.5x in a 496-loop interferometer, while maintaining sensitivity to oscillating signals \cite{supplement}. 

%See End Matter for further details on cost function choice. 

The $N=8$ optimized sequence enables a 50x increase in the loop number. Fig. 2(c) demonstrates that with optimized laser phases, interferometer visibility degrades more gradually than in the unoptimized case as loop number increases. Consequently, interferometer visibility can be resolved beyond $500$ loops compared to the maximum $~10$ loops for the unoptimized sequence under the same imperfect atom-optics pulses.
The red curve is a fundamental visibility limit set by the $21.6 \; \mu$s lifetime of the $^3P_1$ state--our data is within a factor of 4 of this limit \cite{supplement}. We use $80$ ns $\pi$ pulse durations, with $80$ ns of deadtime between pulses.
Further, by increasing the deadtime, we experimentally verify that the benefits of the optimization are preserved when the distance between neighboring atom-laser interactions points (where different sets of trajectories converge (Fig. \ref{fig:resonant_ai_schematic})) is increased beyond a coherence length; we only observe a reduction in visibility consistent with the effects of spontaneous emission as the interferometer duration is increased \cite{supplement}.

We demonstrate that the phase response of the interferometer with the optimized sequence and imperfect mirror pulses is comparably sensitive to resonantly oscillating signals as the expected response of an interferometer with perfect atom-optics pulses at $L=504$ loops. We apply a signal in the form of an oscillating laser phase, $\delta \phi \sin(\omega t_i)$, added to $i$th mirror pulse of the optimized sequence, where the pulse occurs at time $t_i$. An analogous signal could emerge from motion of the atom relative to the laser beam, which results in a change in phase associated with each atom-light interaction \cite{supplement}. 
Moreover, when transforming into a frame that rotates with the oscillating laser phase, the dynamics of the system are equivalent to one in which there are oscillations in the atom's resonance frequency \cite{metcalf1999laser}, e.g. due to dark matter \cite{Arvanitaki2018_darkmatter}.
In Fig. \ref{fig:rank3_introduction}(d), the interferometer phase response to a resonantly oscillating signal is in close agreement with the expected thousandfold phase amplification factor, indicated by the black line.
We leverage this $2L \approx 1000$ phase amplification to resolve the digitization inherent to the AD9910 Direct Digital Synthesizer used to modulate the interferometer beam's phase.
The 16 bit phase resolution yields a minimum modulation amplitude of $\frac{2\pi}{2^{16}} = 96 \; \mu$rad, corresponding to toggling the least significant bit. The inset of panel (d) shows that the error in the measured interferometer phase is smaller than the amplified phase shift from bit flips in a 504-loop sequence, allowing the phase shift from bit flips to be clearly resolved. Fig. \ref{fig:rank3_introduction}(e) shows the narrowing of the resonant response as a function of signal frequency as loop number is increased, consistent with the theoretical response function of an interferometer with perfect atom-optics operations \cite{supplement}. Fig. \ref{fig:rank3_introduction}(f) illustrates the experimental apparatus (see End Matter for further apparatus details). The resilience of the optimized sequence to deviations in overall laser detuning and Rabi frequency are discussed in the End Matter.

\begin{figure}[hbt!]
    \centering
    \includegraphics[scale = 1]{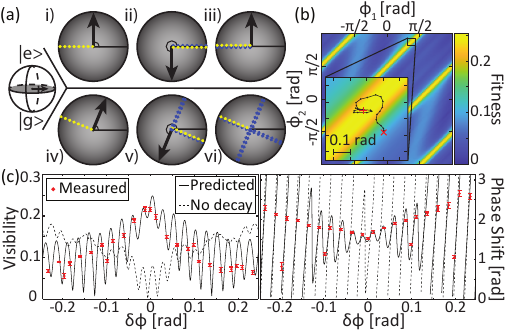}
    \caption{(a) Simplified depiction of a Bloch sphere from a top-down view, composed of only atoms that have undergone spontaneous emission decay. The black arrow indicates the axis of state rotation based on the pulse phase, the yellow (light) represents the semicircular arc formed due to atom decay at various times during the corresponding pulse and subsequent state rotation during the remainder of the pulse, and the blue (dark) represents arcs from previous pulses. (i,ii,iii) Bloch sphere after the first, second, and third mirror pulses, respectively, in the alternating $\pm \pi /2$ phase sequence, showing a non-zero phase formed cumulatively from decayed atoms. (iv,v,vi) Bloch sphere after the first, forth, and eighth mirror pulses, respectively, in the optimized phase sequence (see Fig. \ref{fig:rank3_introduction}), which achieves a symmetric distribution of decayed atoms (arrow removed for visual clarity in vi)). (b) Simulation results of fitness values for pulse sequences over the $\phi_1$-$\phi_2$ parameter space. The inset details an example optimization trajectory for a closed-loop search method (``$\times$": starting point, ``$+$": open-loop optimization result) (see End Matter for further details). (c) Visibility and phase shift for the alternating $\pm \pi /2$ phase sequence over 128 loops vs. amplitude $\delta \phi$ of a resonantly oscillating signal. For $\left|\delta \phi\right| \lesssim 0.1$, visibility markedly increases, but the phase shift shows a largely insensitive response.   For example, when shifting from $\delta \phi = 0$ to the neighboring data point with $\delta \phi = \pi/256$, the phase shift for a fully phase sensitive sequence--such as our optimized sequence--would change by approximately $\pi$.  Here, the phase shift only changes by $\mathcal{O}(100\;\text{mrad})$.  Data are consistent with simulations that include spontaneous emission decay (solid curves) and diverge from those excluding it (dashed curves).}
    \label{fig:cumulative_phase}
\end{figure}

\textit{Spurious interference from spontaneous emission.} Spontaneous emission occurs stochastically throughout the interferometer sequence \cite{chrostoski2024error}, causing atoms to lose all prior phase information stored between the ground and excited states upon decay.
Spontaneous emission is commonly treated as contributing to an incoherent background in atom interferometry \cite{kovachy2015_halfmeter}, though it has been noted that spontaneous emission does not always destroy atom interference \cite{dubetskii1985interference,dubetsky1999matter}. However, atoms undergoing spontaneous emission during a pulse can re-emerge into a coherent superposition due to the remaining pulse duration, 
with the phase of the laser imprinted onto the phase of the transferred population.
We find that certain repeating phase sequences can synchronize the re-emerging atomic coherence from one pulse with that from previous pulses, leading to the build-up of an interference signal with a well-defined cumulative phase that is largely insensitive to oscillating signals of interest. At high loop numbers, this spurious interference pattern sometimes becomes dominant, obscuring the intended signal (Fig. \ref{fig:cumulative_phase}). Furthermore, we find that imposing specific constraints on the pulse sequence can suppress this spurious interference by making the final phase distribution of the decayed atoms average to zero--for example, groups of $N=8$ mirror pulses with the pattern $\phi_1,\phi_2,\phi_2+\pi,\phi_1+\pi,\phi_1+\pi,\phi_2+\pi,\phi_2,\phi_1$ for arbitrary phases $\phi_1$ and $\phi_2$.  We note that the open-loop optimized sequence obeys this constraint.  Fig.  \ref{fig:cumulative_phase} depicts both a phase sequence that leads to cumulative phase buildup involving mirror pulses with phase alternating between $\pm \pi/2$ \cite{bishof2013optical}, and another where we have successfully mitigated this effect using the above constraint. To conceptually illustrate cumulative phase build-up, we consider the interferometer in momentum space so that we can interpret our measurement as an average over atoms with a distribution of momenta $p$, with each atom acting as a two-level system corresponding to the coupling of states $\ket{^1S_0,p}$ and $\ket{^3P_1,p+\hbar k}$.  This allows a Bloch sphere \cite{metcalf1999laser} visualization and optical Bloch equation (OBE) simulations \cite{supplement}.

\textit{Outlook.} Our general optimization framework is not specific to any particular interferometer geometry or atom optics type, thus opening the new avenue of adapting it to enhance various classes of atom interferometers, and matter-wave interferometers more generally. In future research, we plan to apply our optimization method to interferometers utilizing different atom optics and geometries, including those employing broadband dynamical decoupling \cite{zaheer2023quantum}. We also aim to explore both single and multiloop interferometer configurations with larger momentum splittings, where the additional momentum enhances sensitivity \cite{Graham2016_GW}. We have already experimentally observed the adaptability of our open-loop optimization to improve robustness in the latter case, which will be discussed in a future manuscript.

\begin{acknowledgements}
\textit{Acknowledgements.} We thank Andre Carvalho, Garrett Louie, and Viktor Perunicic for valuable discussions, and Garrett Louie also for contributions to the apparatus.  This material is based upon work supported by the U.S. Department of Energy, Office of Science, National Quantum Information Science Research Centers, Superconducting Quantum Materials and Systems Center (SQMS) under contract number DE-AC02-07CH11359.  This work is funded in part by the Gordon and Betty Moore Foundation (Grant GBMF7945), the David and Lucile Packard Foundation (Fellowship for Science and Engineering), the Office of Naval Research (Grant Number N00014-19-1-2181), and the National Institute of
Standards and Technology (Grant Number 60NANB19D168). This research was also supported in part through the computational resources and staff contributions provided for the Quest high performance computing facility at Northwestern University which is jointly supported by the Office of the Provost, the Office for Research, and Northwestern University Information Technology. JG acknowledges support from a National Science Foundation (NSF) Quantum Information Science and Engineering Network (QISE-NET) Graduate Fellowship, funded by NSF award No. DMR-1747426.
\end{acknowledgements}

\textit{Author Contributions.} The author list is grouped into authors who led one or more areas of work (names marked by a \@{*}), followed by additional authors, followed by the principal investigator at the end.  Authors within each group are listed in reverse alphabetical order. TD led apparatus design and construction.  KD, JG, KJ, SS, and YW contributed equally to data collection and analysis. KD and YW led development of the closed-loop optimization approach and spontaneous emission simulations and studies, the latter with supporting contributions from JG.  JG (lead), ZC (supporting), KJ (supporting), and SS (supporting) developed the open-loop optimization approach.  KD (equal), NS (equal), YW (equal), JG (supporting), KJ (supporting), and SS (supporting) commissioned the cold atom source and atom interferometer. TK (lead) and TD (supporting) supervised the work. All authors contributed to conceptualization of the work and validation of the results. KD (equal), JG (equal), KJ (equal), TK (equal), SS (equal), YW (equal), ZC (supporting),  TD (supporting), and NS (supporting) prepared and reviewed the manuscript.

\textit{End Matter on Experimental Apparatus.}---A horizontal laser beam \cite{derose2023high} tuned to the 689\;nm $^1S_0-^3P_1(m_j=0)$ intercombination line performs interferometry on a cloud of $^{88}$Sr atoms released from a  3D magneto-optical trap. The intermediate 21.6\;$\mu$s lifetime of this transition is long enough to be suitable for clock atom interferometry and enables high (MHz-scale) Rabi frequencies with Watt-scale interferometer beams \cite{rudolf2020_689LMT,wilkasonFloquet}, which is advantageous for measuring up to MHz-scale signals, such as in dark matter searches \cite{antypas2022new}. After an interferometer sequence, a 461\;nm push beam spatially separates the ground state from the excited state; then the two populations are read out via fluorescence detection \cite{rudolf2020_689LMT,wilkasonFloquet}.
The interferometer fringe visibility ($v$) and phase ($\Delta\phi$) are extracted by fitting the normalized excited state population ($P_{e}$) as a function of the final beamsplitter phase ($\phi_{B}$), to the expression $\frac{1}{2} (1 + v  \cos[\Delta\phi + \phi_{B}])$.
An M-Labs ARTIQ timing system controls the power, phase, and duration of each pulse via an acousto-optic modulator. The ratio of the laser beam waist to the atom cloud size ($\approx 3$) and the ratio of the Doppler detuning spread to Rabi frequency ($\approx 0.1$, corresponding to a $\approx 4$\;mK temperature) used in this work \cite{supplement} are representative of the ratios anticipated for MAGIS-100. 

For the $<200\;\mu$s sequence durations considered in this paper, the transverse displacement of the atoms in the beam during an experiment cycle due to free fall under gravity is small ($<200$\;nm) compared to the waist of the interferometer beam (mm scale). For longer sequence durations, the interferometer beam can be oriented to be parallel to the direction of gravity to keep the interferometer centered on the atoms during an experiment cycle. 

\textit{End Matter on Open-Loop Optimization Cost Function and Visualization.}---The cost function can be tuned to adjust how strongly the population in stray trajectories is penalized as a function of distance from the central arms. Limiting the spread of trajectories could be useful for accommodating constraints on the interferometer size, like in cases where the interferometer needs to fit inside a satellite \cite{Graham2016_GW} or a magnetically shielded region \cite{Graham2018axion}. The initial exploration of cost function optimization was done in part using the Q-CTRL Boulder Opal package \cite{ball2021software}.  Figure \ref{fig:nl_comparison}(b) presents a visualization of the spatial spread of the atoms for the UR-N sequences at large N, illustrating why these sequences do not perform well per our cost function.  Figure \ref{fig:nl_comparison}(a) shows the corresponding reduction in phase sensitivity to resonantly oscillating signals for these sequences.

\begin{figure}
\includegraphics[scale = 0.96]{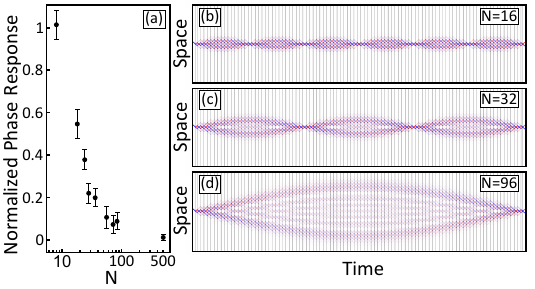}
\caption{\label{fig:nl_comparison}
Comparison of UR-$N$ sequences \cite{Genov_2017} for different values of $N$. (a) Experimentally determined phase sensitivities for a 504-loop sequence. The normalized phase response is $(\Delta\phi[\delta\phi = \pi/(504\times2)] - \Delta\phi[\delta\phi = 0]) / \pi$, where $\Delta\phi$ is the measured phase and $\delta\phi$ the amplitude of resonantly modulating phase. Error bars reflect uncertainties from sinusoidal fits. Panels (b), (c), and (d) show the spread in atom population for $N=16$, $N=32$, and $N=96$ in a 96-loop sequence, as determined by the semi-classical model.
}
\end{figure}

\textit{End Matter on Resilience Against Detuning and Rabi Frequency Errors.}---In Fig. \ref{fig:rank3_robustness}, we examine the resilience of the optimized sequence to deviations in overall laser detuning and Rabi frequency by deliberately introducing errors in these parameters on the order of 10\% of the nominal Rabi frequency. Despite these variations, the interferometer phase remains robust within approximately $\pm100$ mrad and the visibility remains above 50\% of its peak value.

\begin{figure}
\includegraphics[scale = 0.33]{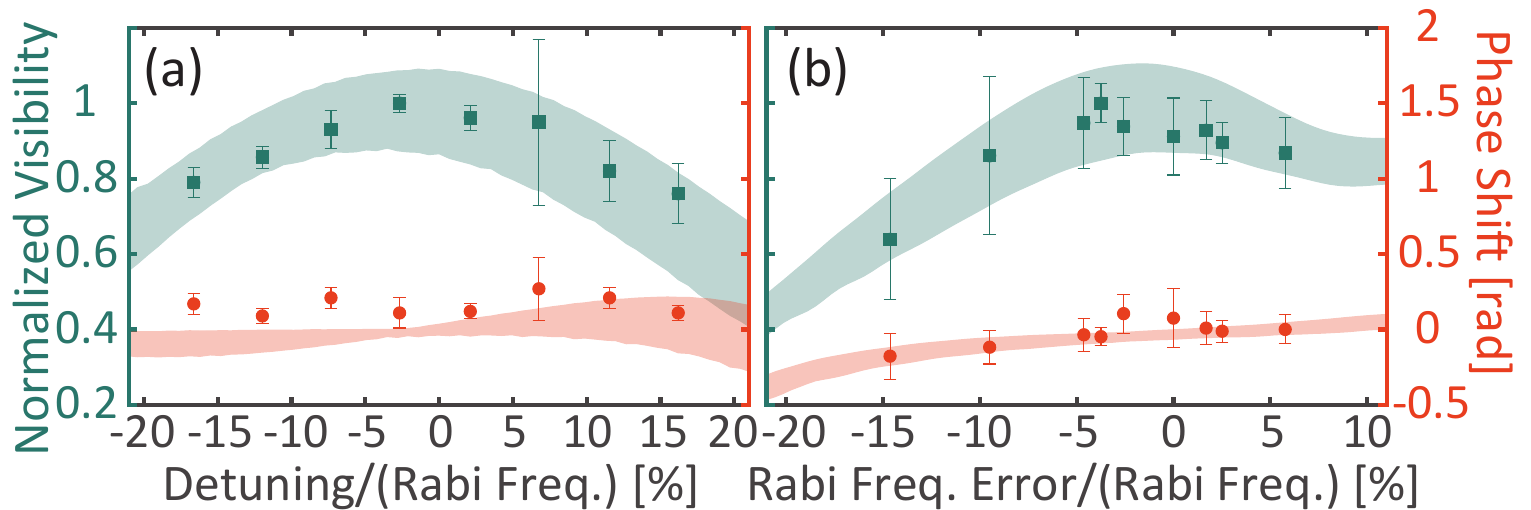}
\caption{\label{fig:rank3_robustness}
Susceptibility of the $N=8$ optimized sequence to detuning (a) and Rabi frequency error (b) at 504 loops, as determined by experiment (solid points), and simulation (shaded bands, width indicating uncertainty). The interferometer visibilities (blue squares for experiment, upper shaded band for simulation) are normalized to their maximum value, and the interferometer phase is displayed in red circles with lower shaded band for simulation. Each data point is the average obtained from 5 individual fringe scans with the error bar being the standard deviation. The detuning is plotted relative to the estimated  Rabi frequency for atoms at the center of the cloud \cite{supplement}.
}
\end{figure}

\textit{End Matter on Closed-Loop Optimization.}---In order to address experimental errors beyond effects captured by simulations, we survey the $\phi_1$-$\phi_2$ parameter space under the constraint described above by implementing closed-loop optimization, where optimization steps are determined by real-time experimental data \cite{Feng2018,rosi2013fast,weidner2018experimental,chih2021reinforcement,ledesma2023machine}. We maximize a fitness function defined as $f = v \cos^4\left(\Delta\phi_0/2\right)$, where $v$ is the measured visibility and $\Delta\phi_0$ is the measured phase shift without an applied signal \cite{supplement}. We begin the optimization near a point that our OBE simulations suggests maximizes the fitness function. The optimization trajectory converges to a pulse sequence consistent with simulation results and with the results of the open-loop optimization (Fig. \ref{fig:cumulative_phase}(b)).

\end{document}